\documentclass[thmsa]{article}
\usepackage{sw20lart}

\input{tcilatex}
\begin{document}

\author{S. Manoff}
\title{$(\overline{L}_n,g)$-spaces. Length of a vector and angle between two vectors}
\date{e-mail address: smanov@inrne.bas.bg }

\begin{abstract}
The notions length of a vector field and cosine of the angle between two
vector fields are determined by means of the metrics and the corresponding
vector fields over a differentiable manifold with contravariant and
covariant affine connections and metrics [$(\overline{L}_n,g)$-spaces]. The
change of the length of a vector field and the cosine of the angle between
two vector fields along a contravariant vector field are found.
\end{abstract}

\maketitle

\section{Introduction}

In previous papers \cite{Manoff-0}, \cite{Manoff-1}  the notions of
contravariant and covariant affine connections are considered for
contravariant and covariant tensor fields over differentiable manifolds. It
was shown that these two different (not only by sign) connections can be
introduced by means of changing the canonical definition of the bases of
dual vector spaces (respectively of dual vector fields). The deviation
operator and its applications to deviation equations over differentiable
manifolds with contravariant and covariant affine connections [the s.c. ($%
\overline{L}_n,g$)-spaces] are investigated.

In the Einstein theory of gravitation (ETG) kinematic notions related to the
notion relative velocity such as shear velocity tensor (shear velocity,
shear) $\sigma $, rotation velocity tensor (rotation velocity, rotation) $%
\omega $ and expansion velocity (expansion) $\theta $, are used in finding
solutions of special types of Einstein's field equations and in the
description of the properties of the (pseudo) Riemannian spaces without
torsion ($V_n$-spaces). By means of these notions a classification of $V_n$%
-spaces, admitting special types of geodesic vector fields, has been
proposed \cite{Ehlers}. The same kinematic characteristics are also
necessary for description of the projections of the Riemannian (curvature)
tensor and the Ricci tensor along a non-isotropic (non-null) vector field 
\cite{Kramer}, and in obtaining and using the Raychaudhuri identity \cite
{Raychaudhuri}, \cite{Hawking} in $V_n$-spaces.

The kinematic characteristics connected with the notion relative velocity
can be generalized for vector fields over differentiable manifolds with
contravariant and covariant affine connections and metrics [($\overline{L}%
_n,g$)-spaces] so that in the case of ($L_n,g$)- and $V_n$-spaces [as a
special case of ($\overline{L}_n,g$)-spaces], and for normalized
non-isotropic vector fields these characteristics are the same as those
introduced in the ETG. In an analogous way as in the case of the kinematic
characteristics, related to the notion of relative velocity, it is possible
to introduce kinematic characteristics, related to the notion of relative
acceleration such as shear acceleration tensor (shear acceleration),
rotation acceleration tensor (rotation acceleration) and expansion
acceleration \cite{Manoff-2}.

The corresponding for ($\overline{L}_n,g$)-spaces notions of relative
velocity and relative acceleration are considered in \cite{Manoff-1}. By
means of these kinematic characteristics, several other types of notions
such as shear velocity, shear acceleration, rotation velocity, rotation
acceleration, expansion velocity and expansion acceleration are
investigated. The connections between the kinematic characteristics related
to the relative acceleration and these related to the relative velocity are
also found. The auto-parallel vector fields in ($\overline{L}_n,g$)-spaces
are classified on the basis of the kinematic characteristics. The
generalizations compared with those in ($L_n,g$)-spaces (differentiable
manifold with affine connection and metric) appear only in the explicit
forms of the expressions, written in a corresponding basis (or in other
words - only in index forms).

In this paper the basic notions of the length of a vector field and the
cosine of the angle between two vector fields are considered over $(%
\overline{L}_n,g)$-spaces as well as their changes along an other vector
field. These notions are necessary for introduction and consideration of
different types of transports in $(\overline{L}_n,g)$-spaces such as
Fermi-Walker transports \cite{Manoff-3} and conformal transports \cite
{Manoff-4}.

\section{Length of a contravariant vector field}

The \textit{square of the length of a contravariant vector field} $u$ is
determined by means of the covariant metric tensor $g$ as 
\begin{equation}
u^2=\pm \mid u\mid ^2=g(u,u)\text{ , }u\in T(M)\text{ , }\mid u\mid \geq 0%
\text{ .}  \label{Ch 7 l.1}
\end{equation}

\begin{definition}
\textbf{\ }\textit{The length of a contravariant vector field }$u$ is the
positive square root of the absolute value of the square of the length of
this field, i.e. 
\begin{equation}
l_u=\,\mid u\mid \,=\,\mid g(u,u)\mid ^{\frac 12}\text{ , \thinspace
\thinspace \thinspace \thinspace }l_u(x)=\,\mid u_x\mid \,=\,\mid u(x)\mid 
\text{ , }x\in M\text{ ,}  \label{Ch 7 l.2}
\end{equation}
where $l_u$ is the length of the contravariant vector field $u$ and $l_u(x)$
is the length of the contravariant vector $u(x)$ in a point $x\in M$.
\end{definition}

With respect to their lengths, the contravariant vector fields can be
divided in two classes: \textit{null- }or\textit{\ isotropic} vector fields (%
$l_u=0 $) and \textit{non-null} or\textit{\ non-isotropic} vector fields ($%
l_u\neq 0 $). In the cases of a \textit{positive definite covariant metric} $%
g$ (Sgn $g=n$, $\dim M=n$) the isotropic vector field is identically equal
to zero ($u\equiv 0$, $u^\alpha \equiv 0$). In the cases of an \textit{%
indefinite covariant metric }$g$ (Sgn $g<n$ or Sgn $g>-n$, $\dim M=n$) the
isotropic vector field with equal to zero length can have different from
zero components in an arbitrary basis, i.e. it is not identically equal to
zero in the points, where it has been defined.

The changes of the length of a contravariant vector field under the action
of the covariant differential operator is determined on one side, by the
action of the covariant operator on a function (here the length $l_u$) over
the manifold and on the other - by the structure of the length $l_u$ itself
and by the commutation relation between the covariant operator and the
contraction operator 
\begin{equation}
\begin{array}{c}
\nabla _\xi u^2=\pm \nabla _\xi (l_u^2)=\nabla _\xi [g(u,u)]=\xi
[g(u,u)]=\pm \xi (l_u^2)= \\ 
=\pm 2l_u(\xi l_u)=(\nabla _\xi g)(u,u)+2g(\nabla _\xi u,u)\text{ ,}
\end{array}
\text{ }  \label{Ch 7 l.3}
\end{equation}

\noindent from where it follows 
\begin{equation}
\xi l_u=\pm \frac 1{2l_u}[(\nabla _\xi g)(u,u)+2g(\nabla _\xi u,u)]\text{
,\thinspace \thinspace \thinspace \thinspace \thinspace \thinspace }l_u\neq 0%
\text{ .}  \label{Ch 7 l.4}
\end{equation}

In the case of an null (isotropic) contravariant vector field $u$, $g(u,u)=0$%
, there is a relation between the covariant derivative $\nabla _\xi g$ of
the covariant metric tensor $g$ and the covariant derivative $\nabla _\xi u$
of the vector field $u$ 
\begin{equation}
(\nabla _\xi g)(u,u)=-2g(\nabla _\xi u,u)\text{ .}  \label{Ch 7 l.5}
\end{equation}

If the contravariant vector field $u$ is transported parallel along the
contravariant vector field $\xi $, the change of the length of $u$ obeys the
condition 
\begin{equation}
\xi l_u=\pm \frac 1{2l_u}(\nabla _\xi g)(u,u)\text{ , \thinspace \thinspace
\thinspace \thinspace }\nabla _\xi u=0\text{ , \thinspace \thinspace
\thinspace }l_u\neq 0\text{ ,}  \label{Ch 7 l.6}
\end{equation}

\noindent and if $l_u=0$ and $\nabla _\xi u=0$, then the condition $(\nabla
_\xi g)(u,u)=0$ follows for the covariant metric $g$ or for the vector field 
$\xi $.

One of the essential characteristics of the different types of transport is
their influence on the change of the length of a contravariant vector field
due to the action of the covariant differential operator. In the case of
manifolds with affine connections and metric which allow different type of
transports of the covariant metric, there are transports under which the
length of a contravariant vector field does not change.

The change of the length of a contravariant vector field under the action of
the Lie differential operator, i. e. the change of length under
draggings-along, can be described in an analogous way as in the cases of
transports 
\begin{equation}
\begin{array}{c}
\pounds _\xi u^2=\pm \pounds _\xi \mid u\mid ^2=\pounds _\xi [g(u,u)]=\pm
2l_u(\xi l_u)= \\ 
=(\pounds _\xi g)(u,u)+2g(\pounds _\xi u,u)\text{ ,}
\end{array}
\label{Ch 7 l.14}
\end{equation}

\noindent from where it follows 
\begin{equation}
\begin{array}{c}
l_u\neq 0:\xi l_u=\pm \frac 1{2l_u}[(\pounds _\xi g)(u,u)+2g(\pounds _\xi
u,u)]\text{ ,} \\ 
l_u=0:(\pounds _\xi g)(u,u)=-2g(\pounds _\xi u,u)\text{ .}
\end{array}
\label{Ch 7 l.15}
\end{equation}

For parallel dragging of $u$ along $\xi $ ($\pounds _\xi u=0$), the change
of $l_u$ can be written in the form 
\begin{equation}
\begin{array}{c}
l_u\neq 0:\xi l_u=\pm \frac 1{2l_u}(\pounds _\xi g)(u,u)\text{ ,} \\ 
l_u=0:(\pounds _\xi g)(u,u)=0\text{ .}
\end{array}
\label{Ch 7 l.16}
\end{equation}

One of the main characteristics of the different draggings-along of a
covariant metric $g$ is the change of length of a contravariant vector field
under draggings-along. The different types of draggings-along of $g$ induce
different changes of the length of a given contravariant vector field.

The length change of contravariant vector fields under different types of ''%
\textit{transport}'' and ''\textit{draggings-along}'' of the covariant
metric $g$ has been used in mathematical models of physical systems,
described by means of differentiable manifolds with affine connections and
metric. The change of length has also been used for giving certain
properties and characteristics with physical interpretation of these systems.

\section{Cosine of the angle between two contravariant vector fields}

The cosine of the angle between two contravariant vector fields is
determined by means of the scalar product of both vector fields and their
lengths. From the relation 
\begin{equation}
g(u,v)=\,\mid u\mid .\mid v\mid .\cos (u,v)=l_u.l_v.\cos (u,v)\text{ ,}
\label{Ch 7 l.22}
\end{equation}

\noindent the definition for the cosine between the vector fields $u$ and $v$
can be written as

\begin{equation}
\cos (u,v)=\frac 1{l_u.l_v}.g(u,v)\text{ , }u,v\in T(M)\text{ , }l_u\neq 0%
\text{ , }l_v\neq 0\text{ .}  \label{Ch 7 l.23}
\end{equation}

From the definition it follows, that two non-isotropic contravariant vector
fields $u$ and $v$ are \textit{orthogonal} to each other, if their scalar
product is equal to zero, i. e. 
\begin{equation}
u\perp v:g(u,v)=0\text{ : }\cos (u,v)=0\text{ ,\thinspace \thinspace
\thinspace \thinspace \thinspace \thinspace \thinspace }u,v\in T(M)\text{ .}
\label{Ch 7 l.24}
\end{equation}

The notion cosine between two isotropic contravariant vector fields or
between one non-isotropic and one isotropic contravariant vector field
cannot be introduced by means of the above definition.

\begin{remark}
If one assumes that (\ref{Ch 7 l.22}) is also fulfilled for the scalar
product of two isotropic (null) vector fields (or for one non-isotropic and
one isotropic vector field), then from (\ref{Ch 7 l.22}) and (\ref{Ch 7 l.24}%
) it could be fixed that: (a) two isotropic contravariant vector fields are
always orthogonal to each other; (b) every isotropic contravariant vector
field is orthogonal to each non-isotropic contravariant vector field. The
last statement leads to the condition $g(u,v)=0$, ($l_u=0$, $l_v\neq 0$),
which is not fulfilled in general. In the case (a) the value of $\cos (u,v)$
as a value of a free determined limited function can be fixed to zero, i. e. 
$\cos (u,v)=0$, $l_u=0$, $l_v=0$. This value corresponds to the notion
orthogonal contravariant non-isotropic (non-null) vector fields. In this
way, \textit{by definition}, the cosine between two isotropic vector fields
is equal to zero. On the other side, every isotropic contravariant vector
field can be presented as a co-linear to other given isotropic vector field,
i.e. if $u\in T(M)$, $l_u=0$, $\Rightarrow \exists v\in T(M)$, $l_v=0$ with $%
u=\kappa .v$, $\kappa \in C^r(M)$. Therefore, under the introduced notion of
orthogonality of two isotropic contravariant vector fields the following
statement is valid: every two co-linear isotropic contravariant vector
fields are orthogonal to each other.
\end{remark}

The cosine change of the angle between two non-isotropic vector fields under
the action of the covariant differential operator is determined on the
grounds of the definition of the notion cosine and the commutation relation
between the covariant and contraction operator 
\begin{equation}
\begin{array}{c}
\nabla _\xi \cos (u,v)=\xi [\cos (u,v)]=\frac 1{l_u.l_v}[(\nabla _\xi
g)(u,v)+g(\nabla _\xi u,v)+g(u,\nabla _\xi v)]- \\ 
-[\frac 1{l_u}(\xi l_u)+\frac 1{l_v}(\xi l_v)].\cos (u,v)= \\ 
=-[\xi (\log l_u)+\xi (\log l_v)].\cos (u,v)+ \\ 
+\frac 1{l_u.l_v}[(\nabla _\xi g)(u,v)+g(\nabla _\xi u,v)+g(u,\nabla _\xi
v)]= \\ 
=-[\xi (\log (l_u.l_v))].\cos (u,v)+ \\ 
+\frac 1{l_u.l_v}[(\nabla _\xi g)(u,v)+g(\nabla _\xi u,v)+g(u,\nabla _\xi v)]%
\text{ .}
\end{array}
\label{Ch 7 l.25}
\end{equation}

The last expression can also be written in the form 
\begin{equation}
\begin{array}{c}
\xi [\cos (u,v)]=\frac 1{l_u.l_v}[(\nabla _\xi g)(u,v)+g(\nabla _\xi
u,v)+g(u,\nabla _\xi v)]- \\ 
-\frac 12\{\pm \frac 1{l_u^2}[(\nabla _\xi g)(u,u)+2g(\nabla _\xi u,u)]\pm
\frac 1{l_v^2}[(\nabla _\xi g)(v,v)+2g(\nabla _\xi v,v)]\}.\cos (u,v)\text{ .%
}
\end{array}
\label{Ch 7 l.26}
\end{equation}

The conditions for transports of the covariant metric $g$ and the conditions
for transports of the contravariant vector fields as well determined the
change of the cosine of the angle between two contravariant vector fields.

Under the action of the Lie differential operator, the change of the angle
between two contravariant vector fields is determined by the action of this
operator on a function over manifold and by its commutation relation with
the contraction operator 
\begin{equation}
\begin{array}{c}
\xi [\cos (u,v)]=\frac 1{l_u.l_v}[(\pounds _\xi g)(u,v)+g(\pounds _\xi
u,v)+g(u,\pounds _\xi v)]- \\ 
-\,\,\,\,\,\,\,[\frac 1{l_u}(\xi l_u)+\frac 1{l_v}(\xi l_v)].\cos (u,v)= \\ 
=\frac 1{l_ul_v}[(\pounds _\xi g)(u,v)+g(\pounds _\xi u,v)+g(u,\pounds _\xi
v)]- \\ 
-[\xi (\log l_u)+\xi (\log l_v)].\cos (u,v)= \\ 
=\frac 1{l_u.l_v}[(\pounds _\xi g)(u,v)+g(\pounds _\xi u,v)+g(u,\pounds _\xi
v)]- \\ 
-\frac 12\{\pm \frac 1{l_u^2}[(\pounds _\xi g)(u,u)+2g(\pounds _\xi u,u)]\pm
\frac 1{l_v^2}[(\pounds _\xi g)(v,v)+2g(\pounds _\xi v,v)]\}.\cos (u,v)\text{
.}
\end{array}
\label{Ch 7 l.35}
\end{equation}

Different draggings-along determine the change of the cosine of the angle
between two contravariant non-isotropic vector fields dragged along other
contravariant vector field.

The change of the cosine of the angle between two contravariant
non-isotropic vector fields under transport or dragging- along is a
characteristic of great importance for mathematical models, describing
physical systems by means of vector fields over differentiable manifolds
with contravariant and covariant affine connections. The geometric
characteristics of vector fields (length, angle between two vector fields)
have been connected with the characteristics of the physical system and in
this way the kinematic structure can be determined for describing physical
processes.

\section{Length of a covariant vector field}

The square of the length of a covariant vector field is determined as 
\begin{equation}
\overline{g}(p,p)=p^2=\pm \mid p\mid ^2=\pm \,\,l_p^2\text{ , }p\in T^{*}(M)%
\text{ , }l_p\geq 0\text{ .}  \label{Ch 7 l.40}
\end{equation}

By means of this definition the notion length of a covariant vector field is
introduced.

\begin{definition}
\textbf{\ }\textit{The length of a covariant vector field }$p$ is the
positive square root of the absolute value of the square of the length of
this field, i.e. 
\begin{equation}
l_p=\,\mid p\mid \,=\,\mid \overline{g}(p,p)\mid ^{\frac 12}\text{
,\thinspace \thinspace \thinspace }l_p(x)=\,\mid p_x\mid \,=\,\mid p(x)\mid 
\text{ , }x\in M\text{ ,}  \label{Ch 7 l.41}
\end{equation}
where $l_p$ is the length of the covariant vector field $p$ and $l_p(x)$ is
the length of the covariant vector $p(x)$ in a point $x\in M$.
\end{definition}

With respect to their lengths the covariant vector fields are also divided
in two types: \textit{isotropic (null), }$(l_p=0$, $l_p(x)=0$, $x\in M)$,%
\textit{\ }and \textit{non-isotropic (non-null), }$(l_p\neq 0$, $l_p(x)\neq 0
$, $x\in M)$, covariant vector fields and vectors. In the cases of definite
contravariant metric $\overline{g}$, ($Sgn\overline{g}=n$, $\dim M=n$), the
isotropic covariant vector filed is identically equal to zero ($p\equiv 0$, $%
p_\alpha \equiv 0$). In the cases of indefinite contravariant metric $%
\overline{g}$, ($Sgn\overline{g}<n$ or $Sgn\overline{g}>-n$, $\dim M=n$) an
isotropic (null) covariant vector field would have different from zero
components in an arbitrary basis, i. e. it is not equal to zero in the
region where it is defined.

The change of the length of a covariant vector field $p$ under the action of
the covariant differential operator $\nabla _\xi $ can be found in an
analogous way as in the case of contravariant vector field 
\begin{equation}
\xi l_p=\pm \frac 1{2l_p}[(\nabla _\xi \overline{g})(p,p)+2\overline{g}%
(\nabla _\xi p,p)]\text{ , \thinspace \thinspace \thinspace \thinspace
\thinspace }l_p\neq 0\text{ .}  \label{Ch 7 l.42}
\end{equation}

The change of the length of a covariant vector field $p$ under the action of
the Lie differential operator $\pounds _\xi $ is determined by the
expression 
\begin{equation}
\xi l_p=\pm \frac 1{2l_p}[(\pounds _\xi \overline{g})(p,p)+2\overline{g}%
(\nabla _\xi p,p)]\text{ .}  \label{Ch 7 l.43}
\end{equation}

For isotropic (null) covariant vector fields the following relations are
fulfilled 
\begin{equation}
\begin{array}{c}
(\nabla _\xi \overline{g})(p,p)=-2\overline{g}(\nabla _\xi p,p)\text{ ,
\thinspace \thinspace \thinspace \thinspace \thinspace \thinspace \thinspace
\thinspace \thinspace \thinspace }l_p=0\text{ ,} \\ 
(\pounds _\xi \overline{g})(p,p)=-2\overline{g}(\pounds _\xi p,p)\text{ ,
\thinspace \thinspace \thinspace \thinspace \thinspace \thinspace \thinspace
\thinspace \thinspace \thinspace \thinspace \thinspace \thinspace }l_p=0%
\text{ .}
\end{array}
\label{Ch 7 l.44}
\end{equation}

Different transports and draggings along of the contravariant metric tensor $%
\overline{g}$ induce analogous changes of the covariant vector fields as in
the case of contravariant vector fields.

\section{Cosine of the angle between two covariant vector fields}

By means of the scalar product of two covariant vector fields and the length
of a covariant vector field, the cosine of the angle between two covariant
vector fields is determined by the relation 
\begin{equation}
\overline{g}(p,q)=\,\mid p\mid .\mid q\mid .\cos (p,q)=l_p.l_q.\cos (p,q)%
\text{ ,\thinspace \thinspace \thinspace \thinspace \thinspace \thinspace
\thinspace \thinspace }p,q\in T^{*}(M)\text{ .}  \label{Ch 7 l.45}
\end{equation}

\begin{definition}
The cosine of the angle between two non-isotropic covariant vector fields $p$
and $q$ is by definition 
\begin{equation}
\cos (p,q)=\frac 1{l_pl_q}.\overline{g}(p,q)\text{ ,\thinspace \thinspace
\thinspace \thinspace \thinspace \thinspace \thinspace \thinspace \thinspace
\thinspace \thinspace \thinspace }p,q\in T^{*}(M)\text{ ,\thinspace
\thinspace \thinspace }l_p\neq 0\text{ ,\thinspace \thinspace \thinspace
\thinspace }l_q\neq 0\text{ .}  \label{Ch 7 l.46}
\end{equation}
\end{definition}

\begin{definition}
\textbf{\ }Two covariant vector fields $p$ and $q$ are \textit{orthogonal to
each other}, when 
\begin{equation}
\overline{g}(p,q)=0\text{ .}  \label{Ch 7 l.47}
\end{equation}
\end{definition}

If the two covariant vector fields $p$ and $q$ are null (isotropic) vector
fields, then from the last definition (\ref{Ch 7 l.45}) it follows that they
are orthogonal to each other.

The change of the cosine of the angle between two covariant vector fields
under the action of the covariant differential operator $\nabla _\xi $ can
be presented in the form 
\begin{equation}
\begin{array}{c}
\xi [\cos (p,q)]=\frac 1{l_pl_q}[(\nabla _\xi \overline{g})(p,q)+\overline{g}%
(\nabla _\xi p,q)+\overline{g}(p,\nabla _\xi q)]- \\ 
-[\xi (\log l_p)+\xi (\log l_q)].\cos (p,q)= \\ 
=\frac 1{l_pl_q}[(\nabla _\xi \overline{g})(p,q)+\overline{g}(\nabla _\xi
p,q)+\overline{g}(p,\nabla _\xi q)]- \\ 
-\frac 12\{\pm \frac 1{l_p^2}[(\nabla _\xi \overline{g})(p,p)+2\overline{g}%
(\nabla _\xi p,p)]\pm \\ 
\pm \frac 1{l_q^2}[(\nabla _\xi \overline{g})(q,q)+2\overline{g}(\nabla _\xi
q,q)]\}.\cos (p,q)\text{ .}
\end{array}
\label{Ch 7 l.48}
\end{equation}

The change of the cosine of the angle between two covariant vector fields
under the action of the Lie differential operator $\pounds _\xi $ can be
found in the form 
\begin{equation}
\begin{array}{c}
\xi [\cos (p,q)]=\frac 1{l_pl_q}[(\pounds _\xi \overline{g})(p,q)+\overline{g%
}(\pounds _\xi p,q)+\overline{g}(p,\pounds _\xi q)]- \\ 
-[\xi (\log l_p)+\xi (\log l_q)].\cos (p,q)= \\ 
=\frac 1{l_pl_q}[(\pounds _\xi \overline{g})(p,q)+\overline{g}(\pounds _\xi
p,q)+\overline{g}(p,\pounds _\xi q)]- \\ 
-\frac 12\{\pm \frac 1{l_p^2}[(\pounds _\xi \overline{g})(p,p)+2\overline{g}%
(\pounds _\xi p,p)]\pm \\ 
\pm \frac 1{l_q^2}[(\pounds _\xi \overline{g})(q,q)+2\overline{g}(\pounds
_\xi q,q)]\}.\cos (p,q)\text{ .}
\end{array}
\label{Ch 7 l.49}
\end{equation}

The cosine changes of the angle between two non-isotropic covariant vector
fields under transports or draggings-along of the contravariant metric can
be found in analogous way as in the case of the covariant metric.

\section{Relative velocity and change of the length of a contravariant
vector field}

Let we now consider the influence of the kinematic characteristics related
to the relative velocity \cite{Manoff-1} upon the change of the length of a
contravariant vector field.

Let $l_\xi =\,\mid g(\xi ,\xi )\mid ^{\frac 12}$ be the length of a
contravariant vector field $\xi $. The rate of change $ul_\xi $ of $l_\xi $
along a contravariant vector field $u$ can be expressed in the form $\pm
\,2.l_\xi .(ul_\xi )=(\nabla _ug)(\xi ,\xi )+2g(\nabla _u\xi ,\xi )$. By the
use of the projections of $\xi $ and $\nabla _u\xi $ along and orthogonal to 
$u$ (see the chapter about kinematic characteristics and relative velocity)
we can find the relations 
\[
\begin{array}{c}
2g(\nabla _u\xi ,\xi )=2.\frac le.g(\nabla _u\xi ,u)+2g(_{rel}v,\xi _{\perp
}) \text{ ,} \\ 
(\nabla _ug)(\xi ,\xi )=(\nabla _ug)(\xi _{\perp },\xi _{\perp })+2.\frac
le.(\nabla _ug)(\xi _{\perp },u)+\frac{l^2}{e^2}.(\nabla _ug)(u,u)\text{ .}
\end{array}
\]

Then, it follows for $\pm \,2.l_\xi .(ul_\xi )$ the expression 
\begin{equation}
\begin{array}{c}
\pm \,2.l_\xi .(ul_\xi )=(\nabla _ug)(\xi _{\perp },\xi _{\perp })+2.\frac
le.[(\nabla _ug)(\xi _{\perp },u)+g(\nabla _u\xi ,u)]+ \\ 
+\,\frac{l^2}{e^2}.(\nabla _ug)(u,u)+2g(_{rel}v,\xi _{\perp })\text{ ,}
\end{array}
\label{Ch 8 2.34}
\end{equation}

\noindent where 
\begin{equation}
g(_{rel}v,\xi _{\perp })=\frac le.h_u(a,\xi _{\perp })+h_u(\pounds _u\xi
,\xi _{\perp })+d(\xi _{\perp },\xi _{\perp })\text{ ,}  \label{Ch 8 2.35}
\end{equation}
\begin{equation}
d(\xi _{\perp },\xi _{\perp })=\sigma (\xi _{\perp },\xi _{\perp })+\frac
1{n-1}.\theta .l_{\xi _{\perp }}^2\text{ .}  \label{Ch 8 2.36}
\end{equation}

For finding out the last two expressions the following relations have been
used: 
\begin{equation}
g(\overline{g}(h_u)a,\xi _{\perp })=h_u(a,\xi _{\perp })\text{ ,\thinspace
\thinspace \thinspace \thinspace \thinspace \thinspace \thinspace }g(%
\overline{g}(h_u)(\pounds _u\xi ),\xi _{\perp })=h_u(\pounds _u\xi ,\xi
_{\perp })\text{ ,}  \label{Ch 8 2.37}
\end{equation}
\begin{equation}
g(\overline{g}[d(\xi )],\xi _{\perp })=d(\xi _{\perp },\xi _{\perp })\text{
,\thinspace \thinspace \thinspace \thinspace \thinspace \thinspace
\thinspace \thinspace }d(\xi )=d(\xi _{\perp })\text{ .}  \label{Ch 8 2.38}
\end{equation}

\textit{Special case}: $g(u,\xi )=l:=0:\xi =\xi _{\perp }$. 
\begin{equation}
\pm \,2.l_{\xi _{\perp }}.(ul_{\xi _{\perp }})=(\nabla _ug)(\xi _{\perp
},\xi _{\perp })+2g(_{rel}v,\xi _{\perp })\text{ .}  \label{Ch 8 2.39}
\end{equation}

\textit{Special case}: $V_n$-spaces: $\nabla _\eta g=0$ for $\forall \eta
\in T(M)$ ($g_{ij;k}=0$), $g(u,\xi )=l:=0:\xi =\xi _{\perp }$. 
\begin{equation}
\pm \,l_{\xi _{\perp }}.(ul_{\xi _{\perp }})=g(_{rel}v,\xi _{\perp })\text{ .%
}  \label{Ch 8 2.40}
\end{equation}

In $(\overline{L}_n,g)$-spaces as well as in $(L_n,g)$-spaces the covariant
derivative $\nabla _ug$ of the metric tensor field $g$ along $u$ can be
decomposed in its trace free part $^s\nabla _ug$ and its trace part $\frac
1n.Q_u.g$ as 
\[
\nabla _ug=\,^s\nabla _ug+\frac 1n.Q_u.g\text{ ,\thinspace \thinspace
\thinspace \thinspace \thinspace \thinspace \thinspace \thinspace \thinspace 
}\dim M=n\text{ ,} 
\]

\noindent where 
\[
\overline{g}[^s\nabla _ug]=0\text{ ,\thinspace \thinspace \thinspace
\thinspace \thinspace \thinspace \thinspace }Q_u=\overline{g}[\nabla _ug]=g^{%
\overline{k}\overline{l}}.g_{kl;j}.u^j=Q_j.u^j\text{ , \thinspace \thinspace
\thinspace \thinspace }Q_j=g^{\overline{k}\overline{l}}.g_{kl;j}\text{ .}
\]

The covariant vector $\overline{Q}=\frac 1n.Q=\frac 1n.Q_j.dx^j=\frac
1n.Q_\alpha .e^\alpha $ is called  \textit{Weyl's covector field}. The
operator $\nabla _u=\,^s\nabla _u+\frac 1n.Q_u$ is called \textit{trace free
covariant operator}.

If we use now the decomposition of $\nabla _ug$ in the expression for $\pm
\,2.l_\xi .(ul_\xi )$ we find the relation 
\begin{equation}
\begin{array}{c}
\pm \,2.l_\xi .(ul_\xi )=(^s\nabla _ug)(\xi ,\xi )+\frac 1n.Q_u.l_\xi
^2+2g(\nabla _u\xi ,\xi )= \\ 
=(^s\nabla _ug)(\xi _{\perp },\xi _{\perp })+ \\ 
+\frac le.[2.(^s\nabla _ug)(\xi _{\perp },u)+2.g(\nabla _u\xi ,u)+\frac
le.(^s\nabla _ug)(u,u)]+ \\ 
+\frac 1n.Q_u.(l_{\xi _{\perp }}^2+\frac{l^2}e)+2.g(_{rel}v,\xi _{\perp })%
\text{ ,}
\end{array}
\label{Ch 8 2.41}
\end{equation}

\noindent where $l_{\xi _{\perp }}^2=g(\xi _{\perp },\xi _{\perp })$, $%
l=g(\xi ,u)$.

For $l_\xi \neq 0:$%
\begin{equation}
ul_\xi =\pm \frac 1{2.l_\xi }(^s\nabla _ug)(\xi ,\xi )\pm \frac
1{2.n}.Q_u.l_\xi \pm \frac 1{l_\xi }.g(\nabla _u\xi ,\xi )\text{ .}
\label{Ch 8 2.42}
\end{equation}

In the case of a parallel transport ($\nabla _u\xi =0$) of $\xi $ along $u$
the change $ul_\xi $ of the length\thinspace \thinspace $l_\xi $ is 
\begin{equation}
ul_\xi =\pm \frac 1{2.l_\xi }(^s\nabla _ug)(\xi ,\xi )\pm \frac
1{2.n}.Q_u.l_\xi \text{ .\thinspace \thinspace \thinspace \thinspace
\thinspace \thinspace \thinspace \thinspace \thinspace \thinspace \thinspace
\thinspace \thinspace \thinspace \thinspace }  \label{Ch 8 2.43}
\end{equation}

\textit{Special case}: $\nabla _u\xi =0$ and $^s\nabla _ug=0$. 
\begin{equation}
ul_\xi =\pm \frac 1{2.n}.Q_u.l_\xi \text{ .\thinspace \thinspace }
\label{Ch 8 2.43a}
\end{equation}

If $u=\frac d{ds}=u^i.\partial _i=(dx^i/ds).\partial _i$, then 
\begin{eqnarray}
l_\xi (s+ds) &\approx &l_\xi (s)+\frac{dl_\xi }{ds}.ds=l_\xi (s)\pm \frac
1{2.n}.Q_u(s).l_\xi (s).ds=  \nonumber  \label{Ch 8 2.43b} \\
&=&(1\pm \frac 1{2.n}.Q_u(s).ds).l_\xi (s)=\triangle _u(s).l_\xi (s)\text{
,\thinspace \thinspace \thinspace \thinspace \thinspace \thinspace
\thinspace }  \nonumber \\
\,\,\,\triangle _u(s) &=&1\pm \frac 1{2.n}.Q_u(s).ds\text{ .}
\label{Ch 8 2.43b}
\end{eqnarray}

Therefore, the rate of change of $l_\xi $ along $u$ is linear to $l_\xi $.

\textit{Special case}: $g(u,\xi )=l:=0:\xi =\xi _{\perp }$. 
\[
\pm 2.l_{\xi _{\perp }}.(ul_{\xi _{\perp }})=(^s\nabla _ug)(\xi _{\perp
},\xi _{\perp })+\frac 1n.Q_u.l_{\xi _{\perp }}^2+2.g(_{rel}v,\xi _{\perp })%
\text{ .} 
\]
\begin{equation}
ul_{\xi _{\perp }}=\pm \frac 1{2.l_{\xi _{\perp }}}.(^s\nabla _ug)(\xi
_{\perp },\xi _{\perp })\pm \frac 1{2n}.Q_u.l_{\xi _{\perp }}\pm \frac
1{l_{\xi _{\perp }}}.g(_{rel}v,\xi _{\perp })\text{ ,\thinspace \thinspace
\thinspace \thinspace \thinspace \thinspace \thinspace \thinspace \thinspace
\thinspace }l_{\xi _{\perp }}\neq 0\text{ .}  \label{Ch 8 2.44}
\end{equation}

\textit{Special case}: Quasi-metric transports: $\nabla _ug:=2.g(u,\eta ).g$%
, \thinspace \thinspace \thinspace $u$, $\eta \in T(M)$. 
\begin{equation}
\pm 2.l_\xi .(ul_\xi )=2.g(u,\eta ).(l_{\xi _{\perp }}^2+\frac{l^2}%
e)+2.[\frac le.g(\nabla _u\xi ,u)+g(_{rel}v,\xi _{\perp })]\text{ .}
\label{Ch 8 2.45}
\end{equation}

\section{Change of the cosine between two contravariant vector fields and
the relative velocity}

The cosine between two contravariant vector fields $\xi $ and $\eta $ has
been defined as $g(\xi ,\eta )=l_\xi .l_\eta .\cos (\xi ,\eta )$. The rate
of change of the cosine along a contravariant vector field $u$ can be found
in the form 
\[
\begin{array}{c}
l_\xi .l_\eta .\{u[\cos (\xi ,\eta )]\}=(\nabla _ug)(\xi ,\eta )+g(\nabla
_u\xi ,\eta )+g(\xi ,\nabla _u\eta )- \\ 
-[l_\eta .(ul_\xi )+l_\xi .(ul_\eta )].\cos (\xi ,\eta )\text{ .}
\end{array}
\]

\textit{Special case}: $\nabla _u\xi =0$, $\nabla _u\eta =0$, $^s\nabla _ug=0
$. 
\[
l_\xi .l_\eta .\{u[\cos (\xi ,\eta )]\}=\frac 1n.Q_u.g(\xi ,\eta )-[l_\eta
.(ul_\xi )+l_\xi .(ul_\eta )].\cos (\xi ,\eta )\text{ .} 
\]

Since $g(\xi ,\eta )=l_\xi .l_\eta .\cos (\xi ,\eta )$, it follows from the
last relation 
\[
l_\xi .l_\eta .\{u[\cos (\xi ,\eta )]\}=\{\frac 1n.Q_u.l_\xi .l_\eta
-[l_\eta .(ul_\xi )+l_\xi .(ul_\eta )]\}.\cos (\xi ,\eta )\text{ .} 
\]

Therefore, if $\cos (\xi ,\eta )=0$ between two parallel transported along $%
u $ vector fields $\xi $ and $\eta $, then the right angle between them
[determined by the condition $\cos (\xi ,\eta )=0$] does not change along
the contravariant vector field $u$. In the cases, when $\cos (\xi ,\eta
)\neq 0$, the rate of change of the cosine of the angle between two vector
fields $\xi $ and $\eta $ is linear to $\cos (\xi ,\eta )$.

By the use of the definitions and the relations: 
\begin{equation}
_{rel}v_\xi :=\overline{g}[h_u(\nabla _u\xi )]=\,_{rel}v\text{ ,\thinspace
\thinspace \thinspace \thinspace \thinspace \thinspace \thinspace \thinspace
\thinspace \thinspace \thinspace \thinspace \thinspace \thinspace \thinspace
\thinspace \thinspace }_{rel}v_\eta :=\overline{g}[h_u(\nabla _u\eta )]\text{
,}  \label{Ch 8 2.46}
\end{equation}
\begin{equation}
\begin{array}{c}
g(\nabla _u\xi ,\eta )=\frac 1e.g(u,\eta ).g(\nabla _u\xi ,u)+g(_{rel}v_\xi
,\eta )\text{ ,} \\ 
g(\nabla _u\eta ,\xi )=\frac 1e.g(u,\xi ).g(\nabla _u\eta ,u)+g(_{rel}v_\eta
,\xi )\text{ ,}
\end{array}
\label{Ch 8 2.47}
\end{equation}
\begin{equation}
(\nabla _ug)(\xi ,\eta )=(^s\nabla _ug)(\xi ,\eta )+\frac 1n.Q_u.g(\xi ,\eta
)\text{ ,}  \label{Ch 8 2.48}
\end{equation}
\begin{equation}
\begin{array}{c}
(^s\nabla _ug)(\xi ,\eta )=(^s\nabla _ug)(\xi _{\perp },\eta _{\perp
})+\frac le.(^s\nabla _ug)(u,\eta _{\perp })+\frac{\overline{l}}e.(^s\nabla
_ug)(\xi _{\perp },u)+ \\ 
+\frac le.\frac{\overline{l}}e.(^s\nabla _ug)(u,u)\text{ ,\thinspace
\thinspace \thinspace \thinspace \thinspace \thinspace \thinspace \thinspace 
}\overline{l}=g(u,\eta )\text{ ,\thinspace \thinspace \thinspace \thinspace
\thinspace \thinspace \thinspace \thinspace }\eta _{\perp }=\overline{g}[%
h_u(\eta )]\text{ ,\thinspace \thinspace \thinspace \thinspace \thinspace }%
l=g(u,\xi )\text{ ,}
\end{array}
\label{Ch 8 2.49}
\end{equation}
\begin{equation}
\begin{array}{c}
(\nabla _ug)(\xi ,\eta )=(^s\nabla _ug)(\xi ,\eta )+\frac 1n.Q_u.g(\xi ,\eta
)= \\ 
=(^s\nabla _ug)(\xi _{\perp },\eta _{\perp })+\frac le.(^s\nabla _ug)(u,\eta
_{\perp })+\frac{\overline{l}}e.(^s\nabla _ug)(\xi _{\perp },u)+ \\ 
+\frac le.\frac{\overline{l}}e.(^s\nabla _ug)(u,u)+\frac 1n.Q_u.[\frac{l.%
\overline{l}}e+g(\xi _{\perp },\eta _{\perp })]\text{ ,}
\end{array}
\label{Ch 8 2.50}
\end{equation}

\noindent the expression of $l_\xi .l_\eta .\{u[\cos (\xi ,\eta )]\}$
follows in the form 
\begin{equation}
\begin{array}{c}
l_\xi .l_\eta .\{u[\cos (\xi ,\eta )]\}=(^s\nabla _ug)(\xi _{\perp },\eta
_{\perp })+\frac le.[(^s\nabla _ug)(u,\eta _{\perp })+g(\nabla _u\eta ,u)]+
\\ 
+\frac{\overline{l}}e.[(^s\nabla _ug)(\xi _{\perp },u)+g(\nabla _u\xi ,u)]+%
\frac{l.\overline{l}}{e^2}.(^s\nabla _ug)(u,u)+ \\ 
+\,\frac 1n.Q_u.[\frac{l.\overline{l}}e+g(\xi _{\perp },\eta _{\perp
})]+g(_{rel}v_\xi ,\eta )+g(_{rel}v_\eta ,\xi )- \\ 
-[l_\eta .(ul_\xi )+l_\xi .(ul_\eta )].\cos (\xi ,\eta )\text{ .}
\end{array}
\label{Ch 8 2.51}
\end{equation}

\textit{Special case}: $g(u,\xi )=l:=0$, $g(u,\eta )=\overline{l}:=0:\xi
=\xi _{\perp }$, $\eta =\eta _{\perp }$. 
\begin{equation}
\begin{array}{c}
l_{\xi _{\perp }}.l_{\eta _{\perp }}.\{u[\cos (\xi _{\perp },\eta _{\perp
})]\}=(^s\nabla _ug)(\xi _{\perp },\eta _{\perp })+\,\frac 1n.Q_u.l_{\xi
_{\perp }}.l_{\eta _{\perp }}.\cos (\xi _{\perp },\eta _{\perp })+ \\ 
+\,g(_{rel}v_{\xi _{\perp }},\eta _{\perp })+g(_{rel}v_{\eta _{\perp }},\xi
_{\perp })-[l_{\eta _{\perp }}.(ul_{\xi _{\perp }})+l_{\xi _{\perp
}}.(ul_{\eta _{\perp }})].\cos (\xi _{\perp },\eta _{\perp })\text{ ,}
\end{array}
\label{Ch 8 2.52}
\end{equation}

\noindent where $g(\xi _{\perp },\eta _{\perp })=l_{\xi _{\perp }}.l_{\eta
_{\perp }}.\cos (\xi _{\perp },\eta _{\perp })$.

The kinematic characteristics related to the relative velocity and used in
considerations of the rate of change of the length of contravariant vector
fields as well as the change of the angle between two contravariant vector
fields could also be useful for description of the motion of physical
systems in $(\overline{L}_n,g)$-spaces.

\begin{center}
\textbf{Acknowledgments}
\end{center}

This work is supported in part by the National Science Foundation of
Bulgaria under Grant No. F-642.


\begin{thebibliography}{9}
\bibitem{Manoff-0}  Manoff S., \textit{Spaces with contravariant and
covariant affine connections and metrics}. Physics of elementary particles
and atomic nucleus (Particles and Nuclei) [Russ. Ed. \textbf{30} (1999) 5,
1211-1269] [Engl. Ed. \textbf{30} (1999) 5, 527-549]

\bibitem{Manoff-1}  Manoff S., \textit{Kinematics of vector fields}. In 
\textit{Complex Structures and Vector Fields}. eds. Dimiev St., Sekigawa K.
(World Sci. Publ., Singapore, 1995), pp. 61-113; \textit{Geodesic and
autoparallel equations over differentiable manifolds}. Intern. J. Mod. Phys. 
\textbf{A 11} (1996) 21, 3849-3874

\bibitem{Ehlers}  Ehlers J., \textit{Beitr\"{a}ge zur relativistischen
Mechanik kontinuierlicher Medien}. Abhandlungen d. Mainzer Akademie d.
Wissenschaften, Math.-Naturwiss. Kl. Nr. \textbf{11 }(1961)

\bibitem{Kramer}  Kramer D., Stephani H., MacCallum M., Herlt E., \textit{%
Exact Solutions of Einstein's Field Equations. }(VEB Deutscher Verlag der
Wissenschaften, Berlin, 1980)

\bibitem{Raychaudhuri}  Raychaudhuri A., \textit{Relativistic cosmology. I}.
Phys. Rev. \textbf{98} (1955) 1123

\bibitem{Hawking}  Hawking S. W., Ellis G., \textit{The Large Scale
Structure of Space-Time} (Cambridge U.P., Cambridge, 1973), \#\# 4.1.,4.4

\bibitem{Manoff-2}  Manoff S., \textit{Kinematics of the relative motion in
the modern gravitational theories}. 13th Intern. Conf. on Gen. Rel. and
Grav., Huerta Grande, Cordoba, Argentina 1992. Contr. Papers. (Cordoba, 1992)

\bibitem{Manoff-3}  Manoff S., \textit{Fermi derivative and Fermi-Walker
transports over }$(\overline{L}_n,g)$\textit{-spaces}. Intern.J. Mod. Phys. 
\textbf{A 13} (1998) 25, 4989-4308; \textit{Fermi derivative and
Fermi-Walker transports over }$(L_n,g)$\textit{-spaces}. Class. and Quantum
Grav. \textbf{15} (1998), 465-477

\bibitem{Manoff-4}  Manoff S.,\textit{\ Conformal transports over }$(L_n,g)$%
\textit{-spaces}. E-print \textbf{gr-qc/990795}; \textit{Conformal
derivative and conformal transports over }$(\overline{L}_n,g)$\textit{%
-spaces. }Intern.J. Mod. Phys. \textbf{A 15} (to appear)
\end{thebibliography}
\end{document}